\documentstyle[12pt]{article}
\begin{document}
\begin{titlepage}
\begin{flushright}
IOP~-~TH~-~97/12\\
{\bf March 1997}
\end{flushright}

\vspace{4.5cm}
\begin{center}
{\Large \bf Natural Flavour-Unifying GUTs: SU(8)~$^\star$}

\vspace{0.5cm}
{Jon Chkareuli}

\vspace{0.6cm}

{\it Institute of Physics, Georgian Academy of Sciences, 380077
Tbilisi, Georgia} \\

\vspace{0.6cm}

\end {center}
\begin{abstract}
Some implications of flavour unification in supersymmetric Grand
Unified Theories are briefly reviewed. We argue that the gauge
hierarchy phenomenon properly interpreted in terms of the natural
missing VEV solution for the adjoint scalar in general $SU(N)$ SUSY
GUT could give insight into both of the basic features of low-energy
physics accommodated in MSSM just as some strict colour-flavour (or
colour-family) interrelation, so the $SU(2)$ symmetry structure of
weak interactions. There, apart from the ordinary triple matter of
MSSM, just three families of the pseudo-Goldstone states appear in
the minimal $SU(8)$ theory to be of necessity detected at a TeV
scale.

\vspace{7cm}

\noindent $\overline{\hspace{6cm}}$ \\
\noindent $^\star${Talk given at  the {\it Trieste Conference on
Quarks and Leptons:  Masses and Mixings},\\ \noindent Trieste,
October 7-11, 1996}
\end{abstract}
\end{titlepage}
\baselineskip=18.3pt

{\large \bf 1 Introduction}
\vspace{0.5cm}

There are still  left several hard problems in Supersymmetric Grand
Unified Theories (SUSY GUTs) suggesting a desired stabilization of
the Higgs masses against the radiative corrections [1].

The main question sounding somewhat notorious addresses, of course,
the soft SUSY breaking itself -- why its scale is called upon to
correlate generally just with the electroweak scale $M_{EW}$ among
the other ones, say, grand unifying scale $M_{GUT}$, the left-right
symmetry scale $M_{LR}$, the family symmetry scale $M_F$ etc might
being in particle physics?  Like any purely geometric symmetry SUSY
could be "blind" to the internal symmetries making no difference
between them.  However, looked just as an opposite case this might
mean some clue towards the gauge hierarchy phenomenon in GUTs, which
if properly interpreted in terms of some stable direction-dependent
"grand" vacuum configuration splitting the ordinary Higgs multiplets
in a particular way, could single out just the electroweak $SU(2)
\otimes U(1)$ symmetry structure surviving down to low energies where
SUSY softly breaks.

Another question of vital importance is that the internal symmetry
breaking caused by scalar supermultiplets of GUT should guarantee
together with a rather peculiar masses of quarks and leptons a nearly
uniform mass spectra for their superpartners with a high degree of
flavour conservation. Presently, among the other possibilities to
alleviate the flavour-changing problem in SUSY GUTs some flavour
symmetry $G_F$ between quark-lepton (squark-slepton) families seems
to be the most natural framework [2].

Thus contemplating the above SUSY GUT issues (gauge hierarchy and
flavour "democracy") one might expect  that there could
appear in the GUT symmetry-broken phase some interrelation between
 both of phenomena principally remaining down to low energies in a
framework of supersymmetric unification. It seems even likely that a
true solution to a doublet-triplet splitting problem in SUSY GUTs (if
 in contrast to a few proposed ones [3-5] was looked for in
general group context) might choose itself the total starting symmetry
of  GUT including some family symmetry $G_F$ provided this were
the case.

Following such a motivation we consider below general SU(N) SUSY
GUTs. We find a stable missing VEV solution which naturally stems
from  general reflection-invariant adjoint scalar superpotential.
Remarkably enough, the solution requires a strict equality of numbers
of  fundamental colours  and flavours  in $SU(N)$ and for the
even-order groups ($N=2n+2$, n=1,2,...) gives in a finite parameter
area the predominant breaking channel
\begin{equation}
 SU(N)\to
SU(n)_C\otimes SU(n)_F\otimes SU(2)_W\otimes U(1)_{1} \otimes
U(1)_{2} \label{1}
\end{equation}
 which definitely favors just  ab
initio three-colour $SU(8)$ case among the other $SU(N)$ GUTs. The
flavour subsymmetry breaking owing to some generic string inspired
extra symmetries of the total Higgs superpotential appears not to
affect markedly the basic adjoint vacuum configuration in the model.
Thus the both salient features of SM just as an interplay between
colours and flavours (or families for due assignment of quarks and
leptons), so the doublet structure of the weak interactions could be
properly understood and well accommodated in the framework of the
minimal $SU(8)$ model.

\vspace{1.5cm}
{\large \bf 2 Missing VEV solutions in SU(N): colour-flavour
interrelation}
\vspace{0.5cm}

We start reminding that a missing VEV
conjecture [6] for doublet-triplet (or $2/(N-2)$ in general)
splitting is that a heavy adjoint scalar $\Phi^{i}_{j}$ $(i,j =1
...N)$ of $SU(N)$ might not develop a VEV in the weak $SU(2)$
direction and through its direct coupling with fundamental chiral
pairs $H_{i}$ and $\bar{H^{i}}$ (containing the ordinary Higgs
doublets) could hierarchically split their masses in the desired
$2/(N-2)$ way. While there was found some special realization of the
missing VEV ans\"{a}tz in $SO(10)$ model [4] the situation in $SU(N)$
theories looks much hopeless. The main obstacle is happened to be a
presence of a cubic term $\Phi^3$ in  general renormalizable Higgs
superpotential $W$ leading to the impracticable trace condition
$Tr\Phi^2 = 0$ for a missing VEV vacuum configuration unless there
occurs the special fine-tuned cancellation between $Tr\Phi^2$ and
driving terms stemming from other parts of $W$ [6].

So, the only way to a natural missing VEV solution in $SU(N)$
theories seems to exclude the cubic term $\Phi^3$ from the
superpotential $W$ imposing some extra reflection symmetry on the
adjoint supermultiplet $\Phi$. While an elimination of the $\Phi^3$
term itself leads usually to the trivial unbroken symmetry case an
inclusion in $W$ another adjoint scalar can, as we show below,
drastically change a situation.

The general renormalizable two-adjoint Higgs superpotential of $\Phi
_s~(s=1,2)$ satisfying the reflection symmetry $\Phi _s \to (-1)^s
\Phi _s$ can be envisioned as
\begin{equation}
W_\Phi
=\frac{M_1}{2}\Phi _1^2+\frac{M_2}{2}\Phi _2^2+ \frac{h}{2}\Phi
_1^2\Phi _2+\frac{\lambda}{3}\Phi _2^3
\label{2}
\end{equation}

The total superpotential apart from the adjoints $\Phi _s$ and the
ordinary Higgs supermultiplets $H$ and $\overline H$ includes also
$N-5$ fundamental chiral  pairs $(\varphi ,
\overline{\varphi})^{(p)}$ $(p=1...N-5)$ which break $SU(N)$ to
$SU(5)$ at some high mass scale (possibly even at Planck mass $M_P$)
\begin{equation}
W=W_{\Phi}+W_H+W_{\varphi}
\label{3}
\end{equation}
We do not consider for a moment the
$W_{\varphi}$ part of the superpotential assuming that some extra
symmetry (see below) makes it possible to ignore its influence on the
formation of the basic adjoint vacuum configurations in the model.

The possible patterns of the adjoints $\Phi _s$ minimizing their own
potential $V_{\Phi}$ can be sought generally in the following
independent diagonal matrix forms $(s=1,2)$

\newpage

\begin{eqnarray}
m_s~~~~~~~~~~~~~~\hspace{0.5cm}
q_s~~~~~~~~~~~~~~
\hspace{0.5cm}
 r_s \hspace{0.5cm}\nonumber\\
\Phi_s=\sigma_sdiag(\overbrace{~1~...~1~}~,
\overbrace{~-{m_s}/{q_s}~...~-{m_s}/{q_s}~}~,
\overbrace{~0~...~0~})
\label{4}
\end{eqnarray} with all
decompositions of $N$, $N=m_s+q_s+r_s$.  A self-consistent
 prerequisite to the supersymmetric minimum following from the
superpotential $W_{\Phi}$ (\ref{2}) through the vanishing $F$-terms
$F_{\Phi_s}=0~(s=1,2)$ for adjoint components in (\ref{4}) implies
 that the spectrum of all basic vacuum configurations in $SU(N)$ is
 bound to include the following four cases only: \\ (i)~The trivial
symmetry-unbroken case ($r_1=N,~r_2=N$),\\ (ii)~The single-adjoint
solutions ($r_1=N,~r_2=0;~m_2\equiv m$)

\def\theequation{5a}
\begin{equation}
\sigma_1=0,~~\sigma_2=-\frac{M_1}{h}ax~~~
(a=\frac{N-m}{N-2m},~x=\frac{M_2}{M_1}\frac{h}{\lambda})~,
\label{5a}
\end{equation}
(iii)~The "parallel" configurations, $\Phi _1\propto \Phi _2$
($r_1=0,~r_2=0;~m_1=m_2\equiv m$)
\def\theequation{5b}
\begin{equation}
\sigma_1=\left(2\frac{\lambda}{h}\right)^{1/2}
\frac{M_1}{h}a(x-1)^{1/2},~~\sigma_2=-
\frac{M_1}{h}a~,
%\eqnum{5b}
\label{5b}
\end{equation}
(iv)~The "orthogonal" configurations, $Tr(\Phi _1\Phi _2)=0$
($r_1=N-2m_1,~r_2=0;~m_1=m_2/2\equiv m/2$)
\def\theequation{5c}
\begin{equation}
\sigma_1=\left(2\frac{\lambda}{h}\right)^{1/2}
\frac{M_1}{h}\left[\frac{N}{N-m}(x-\frac{1}{a})\right]
^{1/2},~~\sigma_2=- \frac{M_1}{h}
\label{5c}
\end{equation}
where the group decomposition number $m$ is generally
different in all the above cases.

So, we can conclude that, while an "ordinary" adjoint $\Phi _2$ having
a cubic term in $W_{\Phi}$ (\ref{2}) develops in all the non-trivial
cases (ii - iv) only a "standard" VEV ($r_2=0$) which breaks the
starting symmetry $SU(N)$ to $SU(m)\otimes SU(N-m)\otimes U(I)$, the
first adjoint $\Phi _1$ can have a new orthogonal solution of
type (\ref{5c}) as well with $SU(N)$ broken along the channel
$SU(m/2)\otimes SU(m/2)\otimes SU(N-m)\otimes U(I)_1\otimes U(I)_2$.
This case corresponds to the missing VEV solutions just we are
looking for if one identifies the $SU(N-m)$ subgroup of $SU(N)$ with
the weak symmetry group while two other $SU(m/2)$ groups must be
identified therewith the groups of  fundamental colours and
flavours, respectively,
\def\theequation{6}
\begin{eqnarray}
N=n_C+n_F+n_W~,~~~~~~~~~~~~~~~~~~~~~~~ \\
n_C=m/2,~~n_F=m/2,~~n_W=N-m \nonumber
%\eqnum{6}
\label{6}
\end{eqnarray}
If so, we are driving at general conclusion that the missing
VEV solutions in $SU(N)$ theories appear only when the numbers of
colours and flavours (or families if one takes a proper
assignment for quarks and leptons under the flavour group
$SU(m/2)_F$) are happened to be equal. As this takes place,
the realistic case with $n_W=2$ corresponding to symmetry breaking
channel (\ref{1}) also occurs for the even-order $SU(N)$ groups
(see Eq.(7)).

Meanwhile the supersymmetric vacuum degeneracy makes no difference
between the different missing VEV solutions (\ref{5c}) as well as
between them in general and the ordinary adjoint ones (\ref{5a},b).
One might expect that supergravity-induced lifting [1] the
vacuum degeneracy modifying the potential $V_{\Phi}$ at the minimum
to (to the lowest order in $k=M_P^{-1}$)
\def\theequation{7}
\begin{eqnarray}
V_{\Phi}\simeq -3k^2|W_{\Phi}|^2
%\eqnum{7}
\label{7}
\end{eqnarray}
would single out a true vacuum configuration in some finite area for
parameters in the superpotential $W_{\Phi}$. Actually
it easily can be shown by direct comparison of all vacua contributions
to $V_{\Phi}$ (\ref{7}) in $O(1/N)$ approximation that the key
missing VEV configuration (\ref{1}) globally dominates over all the
other possible ones in (\ref{5a},b,c) within the following (and in
fact somewhat favorable) parameter area \def\theequation{8}
\begin{eqnarray}
-2\left (\frac{N-2}{N+2}\right
)^{1/3}<x< 2\left (\frac{N-2}{N+2}\right
)^{1/3},~~x=\frac{M_2}{M_1}\frac{h}{\lambda}
%\eqnum{8}
\label{8}
\end{eqnarray}
Remarkably, this dominating vacuum configuration besides the above
general colour-flavour (or colour-family) interrelation provides
somewhat attractive explanation just for a weak $SU(2)$ symmetry
structure coming safely from a grand unifying scale down to low
energies in general $SU(N)$ GUTs.  In essence, there is only one
parameter left for a final specialization of theory -- the number of
colours. It stands to reason that $n_C=3$ and we come to the unique
$SU(8)$ symmetry case with the missing VEV breaking channel
(\ref{1}).

\vspace{1.5cm}
{\large \bf 3 Decoupling of  flavour vacuum configurations:
pseudo-goldstones}
\vspace{0.5cm}

Let us consider now the other parts of the total Higgs superpotential
W (\ref{3}). There $W_H$ is in fact the only reflection-invariant
coupling of the adjoint $\Phi _1$ with a pair of the ordinary
fundamental Higgs supermultiplets $H$ and $\bar{H}$
\def\theequation{9}
\begin{equation}
W_{H}=f\bar{H}\Phi _1 H, ~~~(\Phi _1 \to
-\Phi _1,~~ \bar{H}H\to -\bar{H}H)
%\eqnum{9}
\label{9}
\end{equation}
having the zero VEVs $H=\bar{H}=0$ during the first stage of the
symmetry  breaking. Thereupon $W_H$ turns to the mass term of $H$ and
$\bar{H}$ fields depending on the basic vacuum configurations
(\ref{5a},b,c) in the model. The point is the $\Phi _1$ missing VEV
configuration (\ref{5c}) giving generally heavy masses (of the order
$M_{GUT}$) to them leaves their weak components strictly massless. Thus
there certainly is a natural doublet-triplet splitting in the model
although we drive at the vanishing $\mu$-term on this stage. One can
argue that some $\mu$-term always appears through the radiative
corrections [1] or a non-minimal choice of K\"{a}hler potential
[8] or the high-order terms induced by gravity [7] in the
flavour part of the superpotential $W_{\varphi}$ we are coming to
now.

The flavour symmetry breaking which is also assumed to happen  on the
GUT scale $M_{GUT}$ (not to spoil the standard supersymmetric grand
unification picture) looks in the above favored $SU(8)$ case as
\def\theequation{10}
\begin{equation}
SU(3)_F\otimes U(I)_1\otimes U(I)_2 \to U(1)
%\eqnum{10}
\label{10}
\end{equation}

A question arises: how the missing VEV solution can survive such
a high-scale symmetry breaking (\ref{10}) so as to be subjected at
most to the weak scale order shift?

The simplest way [7,9] could be if there
appeared some generic string-inspired discrete symmetry $Z_k$ which
forbade the mixing between  two sectors $W_{\Phi}+W_{H}$ (I) and
$W_{\varphi}$ (II) in the Higgs superpotential.  Clearly, in a
general SU(N) GUT the accidental global symmetry $SU(N)_I\otimes
SU(N)_{II}$ (and possibly a few $U(1)'$s) being a result of a
$Z$-symmetry must appear and pseudo-Goldstone (PG) states are
produced after the global symmetry breaks. On the other hand
$Z$-symmetries tend to strongly constrain a form of superpotential
$W_{\varphi}$ itself so that contrary to the supersymmetric adjoint
breaking (\ref{1}) the flavour subsymmetry of $SU(N)$ can break
triggered just by the soft SUSY breaking only. Another scenario
[10], suggested recently, uses the anomalous
$U(1)_A$ symmetry which can naturally get untie the sectors (I) and
(II) and induce the high-scale flavour symmetry breaking through the
Fayet-Iliopoulos $D$-term [1]. It stands to reason that both of
ways are fully suited for our case as well to keep the missing VEV
configuration going down to low energies.

\vspace{1.5cm}
{\large \bf 4 Minimal SU(8) model: low-energy predictions}
\vspace{0.5cm}

The time is right to discuss now the particle spectra in the model
concentrating mainly on its favored minimal $SU(8)$ version.

Having considered the basic matter superfields (quarks and leptons
and their superpartners) the question of whether the above flavour
symmetry $SU(3)_F$ is still their family symmetry naturally arises.
Needless to say that among many other possibilities the special
assignment treating the quark-lepton families as the fundamental
triplet of $SU(3)_F$ comes first. Some interesting cases can be found
in Ref.[7,11] showing clearly that the presented $SU(8)$ model
meets a natural conservation of flavour both in the particle and
sparticle sectors, respectively.

Another block, or it would be better to say a superblock of the
low-energy particle spectrum is just three families of PG bosons and
their superpartners of type
\def\theequation{11}
\begin{equation}
5+\bar5~ + ~SU(5)-singlets
%\eqnum{11}
\label{11}
\end{equation}
appearing as a result of breaking of an accidental $SU(8)_I\otimes
SU(8)_{II}$ symmetry in a course of the spontaneous breakdown of the
starting local $SU(8)$ symmetry to SM (\ref{1},\ref{10}) and
acquiring a weak scale order masses due to the soft SUSY breaking and
subsequent radiative corrections or directly through the
gravitational corrections [7]. Normally they are the proper
superpositions of the $SU(5)\otimes SU(3)_F$ fragments $(5,\bar3) +
(\bar5,3)$ in the adjoints $\Phi _s$ of $SU(8)$ and $(5,1)_p+
(\bar5,1)_p$ $(p=1,2,3)$ in its fundamental flavour scalars
$\varphi_p$ and $\bar{\varphi_p}$ (see above).

So, at a low-energy scale one necessarily  has in addition to
three standard families of quarks and leptons (and squarks and
sleptons) just three families of PG bosons and their superpartners
(\ref{11}) which, while beyond the one-loop approximation in the
renormalization group equations (RGEs), will modify the running of
standard gauge $(\alpha_1, \alpha_2, \alpha_S)$ and Yukawa $(Y_t$
primarily, $Y_t={h_t}^2/4\pi)$ couplings from GUT scale $M_{GUT}$ down to
$M_Z$.

We found that the MSSM predictions for $\alpha _S(M_Z)$ and $M_{GUT}$
changed as the PG supermultiplets were included in the RGEs for the
starting large values of top-Yukawa coupling $Y_t$ on the GUT
scale $Y_t(M_{GUT})\geq 0.1$ evolving rapidly towards its infrared fixed
point.

Our results are summarized in Table 1. Performing the
running of the gauge and top-Yukawa couplings from $M_{GUT}$ down to
$M_Z$ (by considering two-loop $\beta$-functions [12]  up-dated
for the $SU(8)$ case with PG supermultiplets) the threshold
corrections related with PG states due to one-loop RGE evolution as
well as the overall one-loop supersymmetric threshold corrections
associated with the decoupling of  supersymmetric particles at
some effective (lumped) scale $T_{SUSY}$ had also been included. We
took for $T_{SUSY}$ the relatively low values $T_{SUSY}=M_Z$ to keep
the sparticle masses typically in a few hundred GeV region [13].

\begin{table}[h]
\caption
{\small{The $SU(8)$ vs MSSM (in square brackets) predictions for
$\alpha _S(M_Z)$, $tan\beta$, $\alpha _{GUT}$ and $M_{GUT}$.
As inputs are taken on the one hand the electroweak scale parameters
$\alpha _{EM}=1/127.9$, $sin^2\theta _W(M_Z)=0.2313$,
$m_t^{pole}=175$ GeV, and on the other hand the GUT scale ones --
some tolerable top-Yukawa coupling and PG mass values on $M_{GUT}$:
$Y_t(M_{GUT})=0.3$ and $\mu _{PG}(M_{GUT})=200$ GeV. For a SUSY scale
$T_{SUSY}=M_Z$ is taken.}}
\begin{center}
\begin{tabular}{llll}\hline  \hline
$\alpha_S(M_Z)$\hspace{1.5cm}& $tan\beta$\hspace{1.5cm}& $\alpha
_{GUT}$\hspace{1.5cm}& $M_{GUT}/10^{16}$ GeV\hspace{1cm}
\\ \hline 0.121 [0.124]\hspace{1.5cm} & 1.12 [1.44]\hspace{1.5cm}
& 0.137 [0.042]\hspace{1.5cm} & 5.08 [2.91]\hspace{1cm}\\
\hline
\hline
\end{tabular}
\end{center} \end{table}

As one can see from Table 1 the $\alpha _S(M_Z)$ value in
 $SU(8)$ model (contrary to MSSM even for the most beneficial,
while still perturbative, value of $Y_t(M_{GUT})=0.3$) is brought into
essentially stable World average value [14] $\alpha _S(M_Z)=0.118\pm
0.003$.

Yet another significant outcome of the $SU(8)$ model turns out to be
very low values of $tan\beta$ if one takes the $Y_t$ fixed-point
solution. As a result the very likely reduction
of the theoretically allowed upper bound on the MSSM lightest Higgs
mass $m_h$ down to $M_Z$ is expected [15]. If so, the Higgs boson
$h$ might be accessible at LEPII.

Meanwhile $b-\tau$ unification, while not generally evolved from the
effective Yukawa couplings of the $SU(8)$ model [7,11], was
found very sensitive to new PG states and, if even were taken on a
GUT scale, actually broke down for any reasonable b-quark mass value
$m_b(m_b)$.

\vspace{1.5cm}
{\large \bf 5 Final remarks}
\vspace{0.5cm}

To conclude any extended $SU(N)$  SUSY GUT containing MSSM appears in
general not to be observationally distinguishable from the standard
supersymmetric $SU(5)$ model [1] provided that its flavour
subsymmetry $SU(N-5)\otimes U(I)$ breaks at some superhigh scale
$M_{N-5}\geq M_5$. However, if its basic adjoint and flavour
vacuum configurations are proved to be essentially untied there
evolve pseudo-Goldstone states (bosons and fermions) at a scale where
supersymmetry softly breaks.  So, one could judge of the starting
symmetry $SU(N)$ and its breaking process to SM by a particular PG
spectrum at low energies.  Depending on the details of their mixing
pattern with the ordinary Higgs sector of MSSM the PG states could
influence appreciably on the particle phenomenology expected at a TeV
scale [16]. Three full-multiplet families of them of type (\ref{11}),
if observed, could say just about $SU(8)$ GUT naturally projected
down to low energies. Otherwise, "between the idea and the reality
...  falls the shadow" [17].

\vspace{2cm}
{\bf Acknowledgments}
\vspace{0.5cm}

It is a pleasure to thank Goran Senjanovi\'{c}, Alexei Smirnov and
their colleagues at ICTP for organizing such a stimulating Trieste
Meeting on flavour physics. I am indebted to Riccardo Barbieri, Zurab
Berezhiani, Barbara Mele, S.Ranjbar-Daemi, Alexei Smirnov and Gena
Volkov for interesting discussions and useful comments and
S.Ranjbar-Daemi also for a warm hospitality at ICTP High Energy
Department where part of this talk was prepared. Many helpful
discussions with my collaborators Ilia Gogoladze and Archil
Kobakhidze are also gratefully acknowledged.

\end{document}